\documentclass[aps,prc,onecolumn,showpacs,amsmath,amssymb,nofootinbib,11pt]{revtex4-2}
\usepackage{graphicx}
\usepackage{color}
\usepackage{times}
\usepackage{inputenc}
\usepackage{bm}
\usepackage{ulem}
\usepackage{multirow}
\usepackage{url}
\usepackage{natbib}
\usepackage{mathrsfs}
\usepackage{physics}
\usepackage{comment}
\usepackage{enumitem}
\usepackage[table,xcdraw]{xcolor}
\usepackage{booktabs,makecell}
\usepackage{comment}
\usepackage{mathrsfs}
\usepackage{float}
\usepackage[colorlinks=true,citecolor=blue,urlcolor=blue,linkcolor=blue]{hyperref}

\usepackage{orcidlink}

\begin{document}

\title{Asteroseismology of Neutron Stars with Axion-Like Particle Mediated Dark Matter}

\author{Pinku Routaray$^{1}$\,\orcidlink{0000-0002-6746-7719}}
\email{routaraypinku@gmail.com}

\author{Rudra N.\ Sahoo$^{1}$}
\email{rudra.sahoo@iopb.res.in}

\author{Bharat Kumar$^{2}$}
\email{kumarbh@nitrkl.ac.in}

\author{M.\ Bhuyan$^{1,3}$}
\email{mrutunjaya.b@iopb.res.in}

\affiliation{\it $^1$ Institute of Physics, Sachivalya Marg, Bhubaneswar-751005, India}
\affiliation{\it $^{2}$ Department of Physics and Astronomy, National Institute of Technology, Rourkela 769008, India}
\affiliation{\it $^{3}$ Homi Bhabha National Institute, Training School Complex, Anushakti Nagar, Mumbai 400094, India}

\date{\today}

\begin{abstract}
We examine the impacts of dark matter (DM) on the observables of the neutron star (NS). The interaction of DM with nucleons intermediate by axion-like particles (ALPs), and the DM-admixed NS (DNS) is formulated in the framework of the relativistic mean field (RMF) formalism. The effects of the DM on the NS are characterized by two parameters: DM Fermi momentum ($q_f$) and DM mass ($m_\chi$). By varying these two parameters, we investigate how the NS's mass, radius, and tidal deformability are altered. Furthermore, we study how the ALP-mediated DM impacts the stability of the NS through the analysis of radial oscillations. In addition, the nonradial oscillation has been investigated by calculating its fundamental mode frequency in the DNS and analyzed how the ALP-mediated DM modifies its behavior.
\end{abstract}

\maketitle

\section{Introduction}  \label{sec:intro}
The dominant portion of the matter in the universe has been classified as DM \cite{Bertone_2005, Bertone_2018, Arbey_Review_DM_2021}. Despite a plethora of findings that provide convincing support for the presence of DM, no findings clarify the particle identification of DM, which continues to be a mystery \cite{Rubin_Galaxy-rotation_1970, Rubin_Galaxy-rotation_1980, Clowe_Coma-cluster_2006, Komatsu_Cosmic-Microwave_2011}. Identifying its fundamental properties has therefore become a major objective in both particle physics and astrophysics. Over the past decades, considerable efforts have been devoted to this problem through direct detection experiments, indirect searches, collider-based studies, and astrophysical observations. Among the available astrophysical systems, NSs provide an exceptional environment for studying DM and its interactions with ordinary matter due their immense gravitational fields and very high densities \cite{Bertone_2008, Bell_2021}. The mass-radius relation, tidal deformability, luminosity, and other macroscopic observables may be affected by the DM component after it has accumulated inside a NS. Such modifications offer an indirect but promising way to investigate the nature of DM through astrophysical observations. Various theoretical frameworks have investigated that in which way the observables of the NSs impacted with DM, including fermionic and bosonic DM models as well as different portal-mediated interactions, each predicting characteristic signatures in the stellar properties. For instance, the accretion of DM in NSs through leptonic targets \cite{Bell_2021_dm_lepton}, dark decay of neutrons \cite{Motta_2018, Husain_2022, Husain_2023, Shirke_2024, Divaris_2026, pinku_pdu_2025}, and various exchange mechanisms between DM and hadronic matter \cite{Kouvaris_2008, Panotopoulos_2017, harish_fmode_2021, Sen_2021, Guha_2021, Dutra_2022, pinku_prd_2023, pinku_mnras_2023, pinku_jcap_2023, pinku_ijmpe_2024, ankit_oscillation_2025, pinku_pdu_2026}. Additionally, the possible existence of of DNS can also be described by the two-fluid approach \cite{Nelson_2019, Ivanytskyi_2020, Davood_2022, Rutherford_2023, Mariani_2023, Diedrichs_2023, Caballero_2024, pinku_prd_2025}. Among these possibilities, axionlike particles (ALPs) have attracted considerable attention. ALPs are light pseudoscalar particles that naturally comes in many extensions of the Standard Model and provide a well motivated framework for mediating interactions between DM and visible matter \cite{Sedrakian_2019, Tanech_alp_2024, Tanech_alp_2025, prashant_alp_2026}.

The RMF formalism provides a well established theoretical framework for describing the properties of dense hadronic matter in NSs and has been widely employed to construct EOS at high densities. Here, the effective nuclear interaction is taken into consideration, along with the meson exchange between nucleons \cite{Hornick_2018, Alford_2022}. However, several alternative approaches have also been developed to describe NS matter, including different microscopic and phenomenological descriptions of hadronic matter and models that allow for the emergence of deconfined quark matter and hadron-quark phase transitions at high densities. These approaches can lead to different predictions for the NS EOS and consequently for observable properties such as the mass-radius relation, tidal deformability, and oscillation modes \cite{Baym_quark_2018, Annala_quark_2020, Ferreira_quark_2020, McLerran_quarkyonic_2019, Zhao_quarkyonic_2020, Cao_quarkyonic_2022}.

The RMF formalism may also be used to simulate the DM's existence within the NS, which allows us to examine its influence on the NS's characteristics \cite{Arpan_2019, pinku_jcap_2023, pinku_ijmpe_2024, ankit_oscillation_2025, Klangburam_2025}. The resulting DNS models must be consistent with terrestrial and cosmological constraints. For example, when an ALP mediates the interaction between DM and nuclear matter (NM), the resulting portal allows DM to be captured by NSs while remaining compatible with terrestrial and cosmological bounds \cite{Klangburam_2025}. The NM-DM interaction inside the NSs is crucial to understand since it effectively governs the oscillation characteristics and stability of the star. Since oscillations are vital for studying the composition of NSs, understanding how DM affects radial as well as nonradial oscillation modes offers a significant advantage in limiting the EOS and DM fingerprints.

Asteroseismology, which examines NS oscillation modes and associated gravitational wave (GW) fingerprints, has evolved into a robust approach for investigating ultra-dense matter. Examining the NS's oscillation will help us gain insights into its intrinsic composition, EOS, and stability criteria \cite{Andersson_1996, Andersson_1998}. Depending upon the kind of perturbation, the NS primarily exhibits two types of oscillation: radial and nonradial. The radial oscillations preserve the spherical symmetry of the star, whereas nonradial oscillations break the spherical symmetry.

 The analysis of radial oscillations will evaluate NS's stability as well as provide important information about its internal structure. Radial oscillation modes are regulated through the linearized perturbation equations of hydrostatic equilibrium, which results in the formation of a corresponding eigenvalue problem \cite{Kokkotas_2001}. Several studies have investigated radial oscillations under different physical scenarios \cite{Panotopoulos_2017, Sagun_2020, pinku_prd_2023, pinku_mnras_2023, Ishfaq_2023, TTSun_2025, ankit_oscillation_2025}. The current work examined the the radial oscillations of NSs with DM which is mediated by ALPs. The existence of DM modifies the foundational equilibrium configuration, potentially altering the critical central density at which instability arises, and whether the NS remains stable or becomes unstable based upon characteristics of the DM and their interaction with NM. 
 
Conversely, the nonradial oscillation perturbation encompasses angular dependence, thereby disrupting the star's spherical symmetry. This oscillation is defined through the spherical harmonic indices, and it is categorized in two types with respect to their corresponding parities, namely, axial (odd-parity) and polar (even-parity) modes. Within the domain of GW ejection, polar modes hold significant relevance because these phenomena pertain to fluid dynamics involving density and pressure fluctuations, resulting in oscillating mass quadrupole moments and consequently effective interaction with spacetime curvature. Moreover, such modes are potentially categorized into different categories based upon characteristics of the restoring force. The primary modes among them involve the fundamental ($f$), pressure ($p$), and gravity ($g$) modes. Of these modes, the $f$-mode is primarily influenced by the star's mean density, rendering it responsive to both the EOS and the overall stellar configuration. The examination of nonradial modes, particularly polar modes, has acquired significant importance with the emergence of GW astronomy. Numerous studies have investigated nonradial oscillations under various physical scenarios \cite{Cowling_1941, Yoshida_1997, Sotani_2011, Flores_2014, Sandoval_2018, Wen_2019, harish_fmode_2021, Pradhan_2022, Zhao_2022, Probit_2024, Flores_2024, Shirke_2024, Probit_prc_2024, Rather_2025, Husain_2026, pinku_pdu_2026}. Given that DM substantially affects the mean average density of NSs, it is expected to modify the nonradial oscillation frequency. This paper employ the general relativistic framework to investigate the fundamental nonradial frequency \cite{Lindblom_1983, Zerilli_1970, Sotani_2021, Zhao_2022, Pradhan_2022, Shirke_2024}.
 
Following is the paper's organizational structure. Sec. \ref{sec:formalism} specifies ALP-mediated DM within the RMF framework and outlines the equilibrium and oscillation formalisms. Sec. \ref{sec:results} presents mass-radius, tidal relations, the radial stability analysis, and the nonradial fundamental mode frequencies. Sec. \ref{sec:conclusions} collects the principal findings and discusses prospects for future studies.

\section{Formalism} \label{sec:formalism}
In this section, we will first specify the construction of DNS EOS. Also, the formalism to calculate the NS observables will be discussed.

\subsection{EOS for DNS}
\label{subsec:eos}
The current study considers the RMF formalism to formulate the hadronic EOS, where the mesons mediate the interaction between nucleons, and also mesons are permitted to demonstrate self-interactions as well as cross-couplings. Here, we took $\sigma$ (isoscalar scalar), $\omega$ (isoscalar vector), and $\rho$ (isovector vector) mesons. The effective Lagrangian density for the NS comprises components from the nucleonic as well as leptonic ($e^-$ and $\mu^-$) domains and may be written as follows \cite{FURNSTAHL_1996, Hornick_2018, Alford_2022}:

\begin{equation}
    \begin{aligned}
        \mathcal{L}_{\rm NS} &= \sum_{\alpha=n,p} \bar{\psi}_{\alpha} \left[ \gamma_{\mu} \left( i\partial^{\mu} - g_{\omega} \omega^{\mu} - \frac{1}{2} g_{\rho} {\tau} \cdot {\rho}^{\mu} \right) \right.
         \left. - \left( M_N -\vphantom{\frac{a}{b}} g_{\sigma} \sigma \right) \right] \psi_{\alpha} + \frac{1}{2}\partial_{\mu}\sigma\partial^{\mu}\sigma - \frac{1}{2}m_{\sigma}^2\sigma^2 \\
        & +\frac{\zeta_0}{4!}g_\omega^2(\omega^{\mu}\omega_{\mu})^2-\frac{\kappa_3}{3!}\frac{g_{\sigma}m_{\sigma}^2\sigma^3}{M_N}-\frac{\kappa_4}{4!}\frac{g_{\sigma}^2m_{\sigma}^2\sigma^4}{M_N^2}
         + \frac{1}{2}m_\omega^2\omega_{\mu}\omega^{\mu} - \frac{1}{4}W^{\mu \nu}W_{\mu\nu} + \frac{1}{2}m_\rho^2{\rho^\mu}\cdot{\rho_\mu} \\
        & - \frac{1}{4}{R^{\mu\nu}}\cdot{R_{\mu\nu}} - \Lambda_\omega g_\omega^2 g_\rho^2 (\omega^\mu \omega_\mu)({\rho^\mu}\cdot{\rho_\mu)} 
         + \sum_{l=e, \mu} \bar{\psi}_k (i\gamma^\mu \partial\mu - m_k) \psi_k
    \end{aligned}
    \label{eq:lagrangian}
\end{equation}

For the above Lagrange density, the energy density can be written as,

\begin{equation}
    \begin{aligned}
        {\cal E}_{NS} = & \sum_{\alpha = n, p} \frac{1}{8\pi^2} \left[k_\alpha E_\alpha^3 + k_\alpha^3 E_\alpha - M^{\star} \ln \left(\frac{k_\alpha + E_\alpha}{M^\star}\right)\right] + \frac{1}{2}m_\sigma ^ 2 \sigma_0^2 + \frac{\kappa_3}{3!} \frac{m_s^2\left(g_\sigma \sigma_0 \right)^3}{g_s^2M_N} \\
        & + \frac{\kappa_4}{4!} \frac{m_s^2\left(g_\sigma \sigma_0 \right)^4}{g_s^2M_N^2} - \frac{1}{2}m_\omega^2 \omega_0^2 - \frac{1}{4!}\frac{\zeta_0\left(g_\omega \omega_0\right)^4}{g_\omega^2} - \frac{1}{2}m_\rho^2 \rho_0^2 - \Lambda_\omega\left(g_\rho g_\omega \rho_0 \omega_0\right)^2 \\ 
        & + \sum_{l = e^-, \mu^-}\frac{1}{\pi^2} \int_{0}^{k_i} k^2\sqrt{k^2 + m_i^2} \,dk 
    \end{aligned}
\end{equation}

where $k_\alpha$ and $E_\alpha = \sqrt{k_\alpha^2 + M^{\star^2}}$ represent the Fermi momentum and Fermi energy of the $\alpha^{th}$ nucleonic species, respectively. Here $M^\star$ defines the effective mass of the nucleon, which is expressed by $M^\star = M_N - g_\sigma \sigma$.

The pressure can be expressed utilizing Gibbs-Duhem relation,
\begin{equation}
    P_{NS} = \sum_{N=\alpha, \ l} \mu_N n_N - {\cal E}_{NS}.
\end{equation}

Chemical potential of the $\alpha^{th}$ nucleonic species $\mu_{\alpha}$ is given by,
\begin{equation}
    \mu_{\alpha} = E_{\alpha} + g_\omega \omega_0 + \frac{1}{2}\tau_{3\alpha}\rho_0 
\end{equation}
where $\tau_{3\alpha}$ is the third component of nucleonic isospin operator.

For leptons, the chemical potential is expressed as,
\begin{equation}
    \mu_l = \sqrt{k_l^2 + m_l^2}.
\end{equation}

In the current study, we have utilized the Hornick2 parameter set in order to compute the hadronic EOS \cite{Hornick_2018}. The parameter set is shown in Table \ref{tab:parameter}. The implementation of the Baym-Pethick-Sutherland (BPS) crust provides a consistent low-density characteristic throughout the EOS \cite{BPS_crust_1971}. Furthermore, at low density, Hornick2 EOS shows a soft nature that is consistent with the bounds illustrated by $\chi$-effective field theory simulations for pure neutron matter \cite{Hornick_2018}. Here, the mass of the nucleon ($M_N$) is considered to be $939$ MeV.

\begin{table*}
\centering
\caption{The measurements of the masses of the mesons $\sigma$, $\omega$, and $\rho$ are expressed in MeV.}
\renewcommand{\tabcolsep}{0.35cm}
\renewcommand{\arraystretch}{1.5}
\scalebox{0.85}{
\begin{tabular}{lcccccccccc}
\hline \hline
Model & $m_\sigma$ & $m_\omega$ & $m_\rho$ & $g_\sigma$   & $g_\omega$   &$g_\rho$   & $\kappa_3$   &$\kappa_4$ & $\Lambda_{\omega}$ & $\zeta_0$\\
\hline
 Hornick2 \cite{Hornick_2018} & $550$ & $783$ & $770$ & $10.993$ & $12.708$  & $10.651$ & $1.6729$ & $-6.6740$ & $0.0272$ & $0$\\
\hline \hline
\end{tabular}}
\label{tab:parameter}
\end{table*}

To incorporate the effects of DM, we have considered the ALP-mediated DM model, where the DM-nucleon interaction is carried by ALPs. Within the confines of such a framework, the standard model (SM) has been extended with a Dirac fermion $\chi$ (the DM candidate) and pseudoscalar ALP $a$. The detailed explanation is in Refs. \cite{Tanech_alp_2024, Tanech_alp_2025, prashant_alp_2026}. Following is the Lagrangian density,

\begin{equation}
    \begin{aligned}
      \mathcal{L}_{\rm DM} & =  \bar{\chi} \big(i\gamma^\mu\partial_\mu-m_\chi \big)\chi + \frac{1}{2}\partial_\mu a\partial^\mu a - \frac{1}{2} m_a^2 a^2 
       + \frac{m_\chi}{f_a}C_\chi \bar{\chi}i\gamma^5 \chi a + \sum_{f}\frac{m_f}{f_a}C_f \bar{\psi}i\gamma^5 \psi a .
    \end{aligned}
    \label{eq:lagrangian_DM}
\end{equation}

Here, $f$ refers to any fermion of the SM type. Specifically, the masses of the fermion, the DM, and the ALP are defined by the parameters $m_f$, $m_\chi$, and $m_a$, respectively. The effective ALP couplings are defined as: 

\begin{equation}
    \begin{aligned}
        g_{aff}=\frac{m_f}{f_a}C_f \ \ \ \ {\rm and} \ \ \ \ g_{a\chi \chi}=\frac{m_\chi}{f_a}C_\chi .
    \end{aligned}
    \label{eq:ALP_couplings}
\end{equation}

The assumption is made that the ALP-fermion couplings $C_f$ are universal for any fermion $m_f$, i.e., $C_f$ and $g_{aff} \propto m_f$. The values of $g_{aff}$ and $g_{a\chi \chi}$ used for this work are $g_{aff} = g_{a\chi \chi} = 10^{-3}$.

The energy density and pressure for the ALP-mediated DM can be written as follows \cite{Klangburam_2025},
\begin{equation}
    \begin{aligned}
        \mathcal{E}_\chi=\frac{1}{\pi^2}\int_0^{q_f} dq \ q^2 \ \sqrt{q^2+m_\chi^2+g_{a\chi\chi}^2a_0^2} \ + \ \frac{1}{2}m_a^2a_0^2,
    \end{aligned}
\end{equation}
\begin{equation}
    \begin{aligned}
        P_\chi=\frac{1}{3\pi^2}\int_0^{q_f} dq \ \frac{q^4}{\sqrt{q^2+m_\chi^2+g_{a\chi\chi}^2a_0^2}} - \ \frac{1}{2}m_a^2a_0^2.
    \end{aligned}
\end{equation}

With the contribution from both NS and DM, the effective EOS for the DNS can be written as $\mathcal{E}=\mathcal{E}_{NS}+\mathcal{E}_\chi$ and $P=P_{NS}+P_\chi$, which will be utilized to calculate the NS properties.

\subsection{Tolman-Oppenheimer-Volkoff Equations} \label{subsec:tov}
For each tabulated EOS, we construct static and spherically symmetric equilibrium NS. The structure is regulated by the Tolman-Oppenheimer-Volkoff (TOV) equations (in units $G = c = 1$) \cite{Oppenheimer_1939, Tolman_1939},

\begin{equation}
\frac{dP}{dr} = -\frac{(\mathcal{E} + P)\bigl[m(r) + 4\pi r^3 P\bigr]}{r\bigl[r - 2m(r)\bigr]},
\label{eq:tov_p}
\end{equation}

\begin{equation}
\frac{dm}{dr} = 4\pi r^2 \mathcal{E}.
\label{eq:tov_m}
\end{equation}

The following boundary conditions could be used to solve above equations:

\begin{itemize}
\item {\it Center} ($r = 0$): $m(r)=0$ and $P(r)=P_c$.
\item {\it Surface} ($r = R$): $m(r)=M$ and $P(r)=0$.
\end{itemize}

In addition to the TOV equations, an additional set of differential equations has been solved to calculate the tidal number $k_2$ for a certain equation of state (EOS). This approach additionally results in another important observable, the dimensionless tidal deformability ($\Lambda$) \cite{Hinderer_2008, Hinderer_2010, Flanagan_2018, Chatziioannou_2020},

\begin{equation}
    \Lambda=\frac{2}{3}k_2\Big(\frac{R}{M}\Big)^5=\frac{2}{5}k_2C^{-5},
    \label{eq:tidal}
\end{equation}
where $C$ is the compactness of the star.

Here, the expression for $k_2$ can be written as, 
\begin{equation}
    \begin{aligned}
        k_2= & \frac{8}{5} C^5(1-2 C)^2\left[2\left(y_2-1\right) C-y_2+2\right] \times\left\{\left[\left(4y_2+4\right) C^4+\left(6 y_2-4\right) C^3-\left(22 y_2-26\right) C^2 \right.\right] \nonumber \\
        & \left.+3\left(5 y_2-8\right) C-3\left(y_2-2\right)\right]2 C  +3(1-2 C)^2 \left.\times\left[2\left(y_2-1\right) C-y_2+2\right] \log (1-2C)\right\}^{-1}\,.
    \end{aligned}
\label{eq:k2}
\end{equation}
 
 $y_2$ is the tidal perturbation which characterize the radial behavior of the induced quadrupolar tidal perturbation inside the NS and it can be determined by solving the following differential equation with initial boundary condition $y(0)=2$,
\begin{equation}
    r\frac{dy_2(r)}{dr} + y_2(r)^2 + y_2(r)F(r) + r^2 Q(r) = 0,
\label{eq:tidal_differential}
\end{equation}
Where,
\begin{equation}
    \begin{aligned}
        F(r) & = \frac{r - 4\pi r^3 \qty{\mathcal{E}(r) - P(r)}}{r - 2M(r)}, \\
        Q(r) & = \frac{4\pi r \qty{5\mathcal{E}(r) + 9P(r) + \frac{\mathcal{E}(r) + P(r)}{\pdv*{P(r)}{\mathcal{E}(r)}}}}{r-2M(r)} - 4 \bqty{\frac{M(r) + 4\pi r^3 P(r)}{r^2\qty{1-2M(r)/r}}}.
    \end{aligned}
\end{equation}

\subsection{Radial Oscillation and Stability} \label{subsec:radial}
On each equilibrium model, we solve for the radial pulsations, which furnish a direct test of dynamical stability. The relativistic perturbation formalism has been employed by conserving the spherical symmetry of the background equilibrium structure \cite{Kokkotas_2001, pinku_prd_2023}.

The radial displacement of an unperturbed system with harmonic time dependence may be written as $\delta r(r,t) = X(r)\, e^{i\omega t}$. Here $X(r)$ defines the amplitude, whereas $\omega$ represents the circular frequency. Linear perturbations then lead to a second-order equation for $X(r)$,
\begin{equation}
    \begin{aligned}
        &c_{s}^{2} X^{\prime \prime}+\left(\left(c_{s}^{2}\right)^{\prime}-Z+ 4 \pi r \gamma P e^{2 \lambda}-\nu^{\prime} \right) X^{\prime} \\
        &+{\left[2\left(\nu^{\prime}\right)^{2}+\frac{2  m}{r^{3}} e^{2 \lambda}-Z^{\prime}-4 \pi (P+\mathcal{E}) Z r e^{2 \lambda} + \omega^{2} e^{2 \lambda-2 \nu} \right]X} =0,
    \end{aligned}
    \label{eq:radial_equation}
\end{equation}
In this context, the prime signifies differentiation with respect to the radial coordinate $r$. $\nu$ and $\lambda$ are the metric functions. $c_s^2 = dP/d\mathcal{E}$ is the square of the speed of sound, and $\gamma = (1+\frac{\mathcal{E}}{P})c_s^2$ is the adiabatic index.

And $Z$ is defined as,
\begin{equation}
    Z(r)=\left(\nu^{\prime}-\frac{2}{r}\right)c_{s}^{2}
    \label{eq:Zerili_function}
\end{equation}

The displacement function could be redefined as,
\begin{eqnarray}
\zeta = r^2e^{-\nu}X \, .
\label{eq:perturbed_equation}
\end{eqnarray}
Using this new variable, Eq. (\ref{eq:radial_equation}) may be reformulated as a Sturm-Liouville differential equation characterized by self-adjoint properties \cite{Kokkotas_2001}
\begin{eqnarray}
\frac{d}{dr}\left( H\frac{d \zeta}{d r}\right)+\left(\omega^{2} W+Q\right) \zeta=0 \, ,
\label{eq:self_adjoint_eq}
\end{eqnarray}
where
\begin{align}
&r^2H=(P+\mathcal{E}) e^{\lambda+3 \nu} c_{s}^{2} \, ,
\nonumber \\
&r^2W=(P+\mathcal{E}) e^{3 \lambda+\nu} \, ,
\nonumber \\
&r^2Q=(P+\mathcal{E}) e^{\lambda+3 \nu}\left((\nu^{\prime})^{2}+\frac{4}{r} \nu^{\prime}- 8 \pi e^{2 \lambda} P\right).
\label{eq:HWQ}
\end{align}
Now, the Lagrangian variation pressure may expresses as, 
\begin{equation}
    \Delta P = -r^{-2} e^{\nu}(P+\mathcal{E}) c_{s}^{2}\zeta^{\prime}.
    \label{eq:pres_variation}
\end{equation}

Physical boundary conditions are:
\begin{itemize}
\item {\it Center} ($r = 0$): The radial oscillations can't displace the fluid element at the center, $X(r=0)=0$.
\item {\it Surface} ($r = R$): The Lagrangian pressure perturbation vanishes, $\Delta P(R) = 0$.
\end{itemize}

The Sturm-Liouville eigenvalue problem possesses a discrete spectrum of real squared frequencies $\omega_n^2$ ($n = 0,1,2,\ldots$) in increasing order. The fundamental mode ($n=0$) has the lowest frequency. A stellar model is dynamically stable against radial perturbations if and only if $\omega_0^2 > 0$; the first instability occurs exactly when $\omega_0^2 = 0$, coinciding with the maximum-mass turning point on the $M(\mathcal{E}_c)$ curve. If $\omega_0^2 < 0$, the star experiences instability characterized by an imaginary frequency at the node, leading to exponential oscillations, preventing a return to its initial form, ultimately resulting in its collapse into a black hole.

For numerical solution, we convert eq. (\ref{eq:radial_equation}) to an equivalent first-order system by defining $\zeta$ and $\eta = H \zeta^\prime$:
\begin{equation}
\frac{d\zeta}{dr} = \frac{\eta}{H}, \qquad
\frac{d\eta}{dr} = -(\omega^2 W + Q)\, \zeta.
\label{eq:radial_1st_order}
\end{equation}

Near the center, a regular Taylor expansion supplies starting values $\eta(0) = 3\zeta(0) H(0)$, with 

\begin{eqnarray}
    H_0=(P(0)+\mathcal{E}(0)) e^{\lambda(0)+3 \nu(0)} c_{s}^{2}(0)
    \label{eq:eta0}
\end{eqnarray}
where $\eta_0$ and $\zeta_0$ are their corresponding values at the center of the star.

By taking $\eta_0 = 1$, we will obtain $\zeta_0 = 1 / (3H_0)$ as the initial value to begin our numerical integration using the shooting method. The values of $\omega$ that fulfill $\eta(r=R)=0$ will provide the necessary radial oscillation modes.

\subsection{Nonradial oscillation} \label{subsec:non_radial}
The nonradial oscillation modes of a spherically symmetric NS can be decomposed into spherical harmonics, $Y_{lm}(\theta,\phi)$, and classified according to their parity. The current study is confined to analysis of the fundamental quadrupolar ($l=2$, $n=0$) polar $f$-mode. This mode corresponds to the lowest-order fluid oscillation, for which pressure acts as the dominant restoring force. Because of its efficient coupling to GWs, the $f$-mode has attracted considerable attention as one of the primary studies of NS asteroseismology.

GW frequencies are derived by addressing an eigenvalue problem grounded in static, spherically symmetric NS models. The frequencies computed are typically complex, where the real component signifies the oscillation frequency of the star ($f$-mode) and the imaginary component denotes the decay rate of these oscillations ($\tau$). These complex frequencies are frequently termed Quasi-Normal Modes (QNM), 
\begin{eqnarray}
    \omega=2\pi f+i\frac{1}{\tau}.
\end{eqnarray}

The relativistic Cowling approximation has previously been used in several works to study the $f$-mode while not taking into account metric perturbation \cite{Pradhan_2021, harishprd_2021, sailesh_2024, pinku_jcap_2023, Probit_2024}. This approximation typically predicts $f$-mode frequencies that are about $10-30\%$ higher than those obtained from a complete general-relativistic treatment \cite{Lindblom_1983,Sotani_2020, Zhao_2022, Pradhan_2022, pinku_pdu_2026}. The present investigation used the general relativistic framework to determine the QNM frequency. Taking into account the relevant boundary conditions analogous to those established by Detweiler and Lindblom \cite{Lindblom_1983}, the perturbation equations were subsequently resolved to obtain the QNM frequency. The regularity criterion has been applied to all perturbation variables at the center of the star. At the center, where $r=0$, the expansion of all requisite variables was successfully performed utilizing Taylor power series. The second boundary condition occurs near the stellar surface, when the pressure fluctuation dissipates. Ultimately, we have established the final boundary condition for the external perturbation of the star, where only outgoing GWs are present at infinity, and the Zerilli equation has been employed to resolve this \cite{Zerilli_1970}. The detailed equations and derivation can be found in the Refs. \cite{Sotani_2021, Zhao_2022, Pradhan_2022, Shirke_2024, pinku_pdu_2026}.

\section{Results and Discussion}  \label{sec:results}
The introduction of ALP-mediated DM modifies EOS, as the dark fermions increase the total energy density while contributing negligibly to the pressure. The degree of EOS softening is primarily governed by $q_f$ and $m_\chi$. To investigate how these DM parameters influence NS's characteristics, we investigate both their equilibrium structure and dynamical behavior. We analyze the mass-radius relation, tidal deformability, and the fundamental frequencies of radial as well as nonradial oscillation modes. The following sections go into depth on how the DM parameters affect these observables.

\subsection{Mass-radius relation}  \label{subsec:mr}
\begin{figure}
    \centering
    \includegraphics[width=1.0\textwidth]{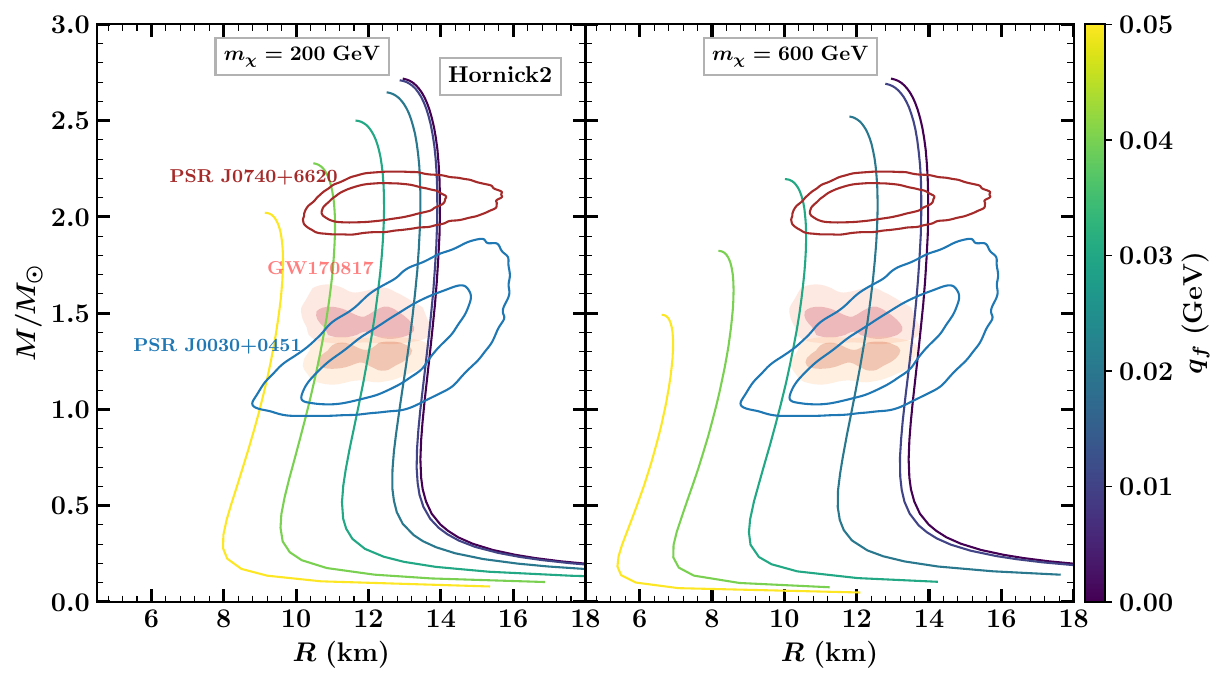}
    \caption{Mass-radius at fixed DM particle mass $m_\chi = 200$ GeV (left) and $m_\chi = 600$ GeV (right), with the $q_f$ varied from $0$ to $0.05$ GeV (color bar). Different observational bounds such as PSR J0030+0451 \cite{Riley_2019}, PSR J0740+6620 \cite{Riley_2019} and GW170817 \cite{Abbott_2017} have been used.}
    \label{fig:mr_mx}
\end{figure}

The mass-radius relationship provides the direct signature of the EOS softening attributed to the ALP-mediated DM. The dark fermions contribute to the total energy density of the star, thereby increasing the energy density for a given baryonic configuration. However, their pressure contribution is comparatively small in the parameter regime considered here. Consequently, a larger fraction of the total energy density is associated with the dark component without a corresponding increase in pressure, resulting in a lower total pressure at a given energy density compared with the pure hadronic case. This reduction in the pressure support effectively softens the EOS and leads to an effect on both the stellar mass and radius. Fig. \ref{fig:mr_mx} represents the effects of DM at different $m_\chi$ with varying $q_f \ (0-0.05 \ {\rm GeV})$. In the left panel, at $m_\chi=200 \ {\rm GeV}$, both mass and radius reduce with increasing $q_f$. The effect is stronger at larger $m_\chi=600 \ {\rm GeV}$ (right panel), as a heavier dark particle carries a higher central energy density with less pressure for the same $q_f$. Observational limitations from several pulsars, including PSR J0740+6620, PSR J0614-3329, and GW data (e.g., GW170817), have been incorporated to ensure coherence between the DNS outcomes and astrophysical data.
\begin{figure}
    \centering
    \includegraphics[width=1.0\textwidth]{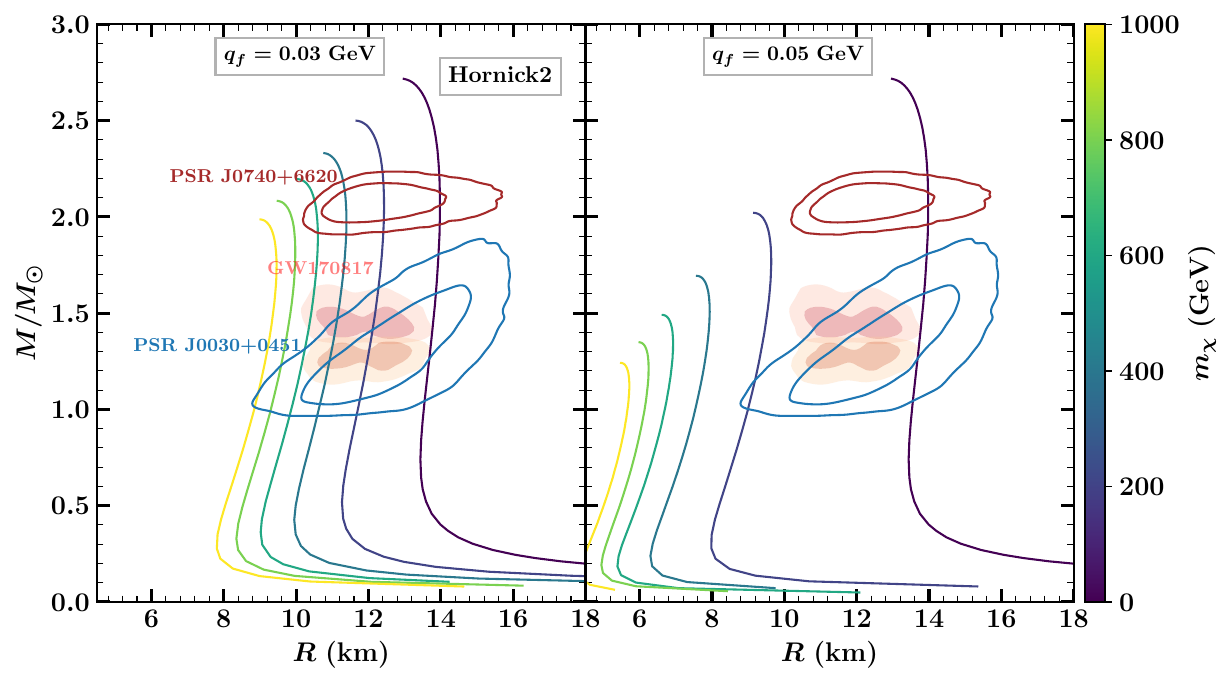}
    \caption{Similar to Fig. \ref{fig:mr_mx}. Fixed DM fermi momentum} $q_f = 0.03$ GeV (left) and $q_f = 0.05$ GeV (right), with the $m_\chi$ varied from $0$ to $1000$ GeV (color bar).
    \label{fig:mr_qf}
\end{figure}

In Fig. \ref{fig:mr_qf}, an analogous trend in the mass-radius relation is also observed while including the DM. In the left panel, it is observed that at constant $q_f = 0.03$ GeV, the maximum mass as well as the corresponding radius decreases with the massive DM particle. This effect is more pronounced in the right panel, where we took $q_f = 0.05$ GeV, and the mass-radius curve failed to satisfy the observational constraints.


\subsection{Dark matter fraction}
Another important metric that provides us insight into the crucial information about the contribution of DM to the structure and observables of NS is the DM fraction within the NS, $f_\chi$. This value defines the effect of DM on the internal structure and total gravitational mass of the DNS. The DM fraction may be calculated using the following equation,
\begin{equation}
    f_\chi = \frac{\displaystyle\int_0^R 4\pi r^2 \mathcal{E}_\chi(r) \ dr}
    {\displaystyle\int_0^R 4\pi r^2 \mathcal{E}(r) \ dr}
    =\frac{M_\chi}{M}.
\end{equation}
The resulting values for the maximum-mass configurations are summarized in Table \ref{tab:DMfraction}. For $m_\chi=200$ GeV, the DM mass fraction increases from $0.47\%$ to $29.20\%$ as $q_f$ increases from $0.01$ to $0.05$ GeV. The associated DM gravitational mass escalates from $0.01 \ M_\odot$ to $0.59 \ M_\odot$, while the maximum total gravitational mass diminishes from $2.70 \ M_\odot$ to $2.02 \ M_\odot$.

For the more massive DM particle with a mass of $600$ GeV, the contribution from DM becomes more substantial. $f_\chi$ escalates from $1.36\%$ at $q_f=0.01$ GeV to $45.16\%$ at $q_f=0.05$ GeV, while the associated DM gravitational mass rises from $0.04 \ M_\odot$ to $0.67 \ M_\odot$. Simultaneously, the upper limit of the total gravitational mass diminishes from $2.68 \ M_\odot$ to $1.49 \ M_\odot$. Consequently, raising either the $q_f$ or $m_\chi$ amplifies the DM component of the overall star mass and results in a more pronounced decrease in the maximum supported gravitational mass.


\begin{table}
\centering
\caption{Maximum total gravitational mass ($M_{\rm max}^{\rm total}$), corresponding DM mass fraction ($f_\chi$), and total DM gravitational mass ($M_\chi$) for different values of the $q_f$ and $m_\chi$. The maximum gravitational mass of the pure hadronic Hornick2 EOS is $2.71 \ M_\odot$.}
\renewcommand{\tabcolsep}{0.35cm}
\renewcommand{\arraystretch}{1.5}
\scalebox{1.0}{
\begin{tabular}{cccc}
\hline
\multicolumn{4}{c}{$m_\chi = 200~\mathrm{GeV}$} \\
\hline
$q_f$ & $M_{\mathrm{max}}^{\mathrm{total}}$ [$M_\odot$] & 
$f_\chi$ [\%] & $M_\chi$ [$M_\odot$] \\
\hline
0.01  & 2.70  & 0.47   & 0.01 \\
0.02  & 2.63  & 3.52   & 0.09 \\
0.03  & 2.49  & 10.34  & 0.26 \\
0.04  & 2.27  & 19.40  & 0.44 \\
0.05  & 2.02  & 29.20  & 0.59 \\
\hline
\multicolumn{4}{c}{$m_\chi = 600~\mathrm{GeV}$} \\
\hline
$q_f$ & $M_{\mathrm{max}}^{\mathrm{total}}$ [$M_\odot$] & 
$f_\chi$ [\%] & $M_\chi$ [$M_\odot$] \\
\hline
0.01  & 2.68  & 1.36   & 0.04 \\
0.02  & 2.51  & 9.34   & 0.23 \\
0.03  & 2.19  & 22.67  & 0.50 \\
0.04  & 1.82  & 35.34  & 0.64 \\
0.05  & 1.49  & 45.16  & 0.67 \\
\hline
\end{tabular}}
\label{tab:DMfraction}
\end{table}

\subsection{Tidal Deformability}  \label{subsec:tidal}
Tidal deformability offers an independent probe for analyzing the internal composition of the NS. Dimensionless tidal deformability ($\Lambda$) may be calculated by solving the tidal perturbation equation together with TOV equations. Because $\Lambda$ depends on both the global compactness and the density profile in the outer layers, it is correlated with. 

\begin{figure}
    \centering
    \includegraphics[width=1.0\textwidth]{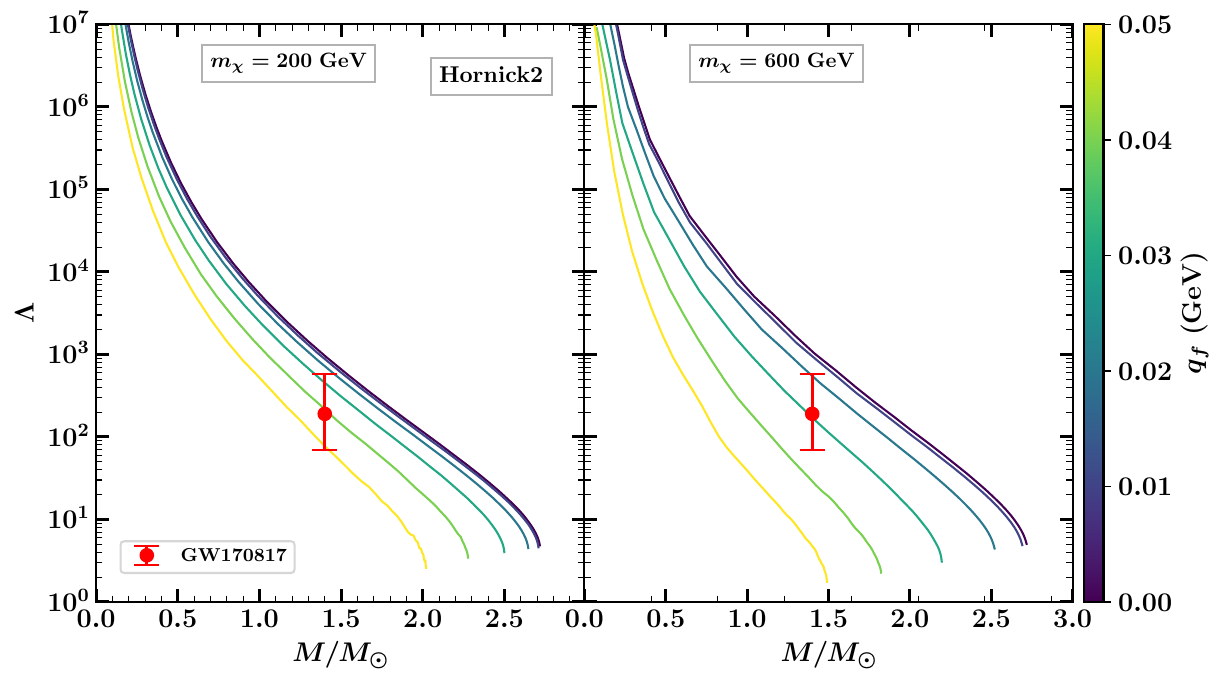}
    \caption{The fluctuation of dimensionless tidal deformability shown with mass of the NS. Left panel: at fixed DM particle mass, $m_\chi = 200$ GeV, the $q_f$ is varied from $0-0.05$ GeV. Similar in the right panel at $m_\chi=600$ GeV. The GW170817 constraint at $1.4 \ M_\odot$ is used in both panels \cite{Abbott_2017}.}
    \label{fig:ml_mx}
\end{figure}

The Fig. \ref{fig:ml_mx} shows the $\Lambda - M$ relation at fixed $m_\chi$, with $q_f$ ranging from $0-0.05$ GeV, which is shown in the color bar. The observational bound on the dimensionless tidal deformability is imposed at canonical mass from the GW170817 event to maintain consistency with the theoretical models. In both panels, it is observed that when we increase the $q_f$, the tidal deformability starts to decrease. This is because the tidal deformability is directly associated with the mass and radius of the star, which shows an effective reduction while including the DM. For the hadronic case, when the $q_f=0$, the canonical tidal deformability ($\Lambda_{1.4}$) exhibit a value $974.61$. At $m_\chi=200$ GeV (left panel), when the $q_f$ increases, the tidal deformability reduces and seems to fall within the range of observational constraint. Similarly, for heavier DM particle mass, at $m_\chi=600$ GeV, the tidal deformability seems to deviate from the observational constraint after $q_f=0.03$ GeV. The mechanism is similar to radius reduction. Thus, it can be state that the radius reduction for larger $q_f$ or $m_\chi$ produces a more compact configuration whose tidal response is weaker.
\begin{figure}
    \centering
    \includegraphics[width=1.0\textwidth]{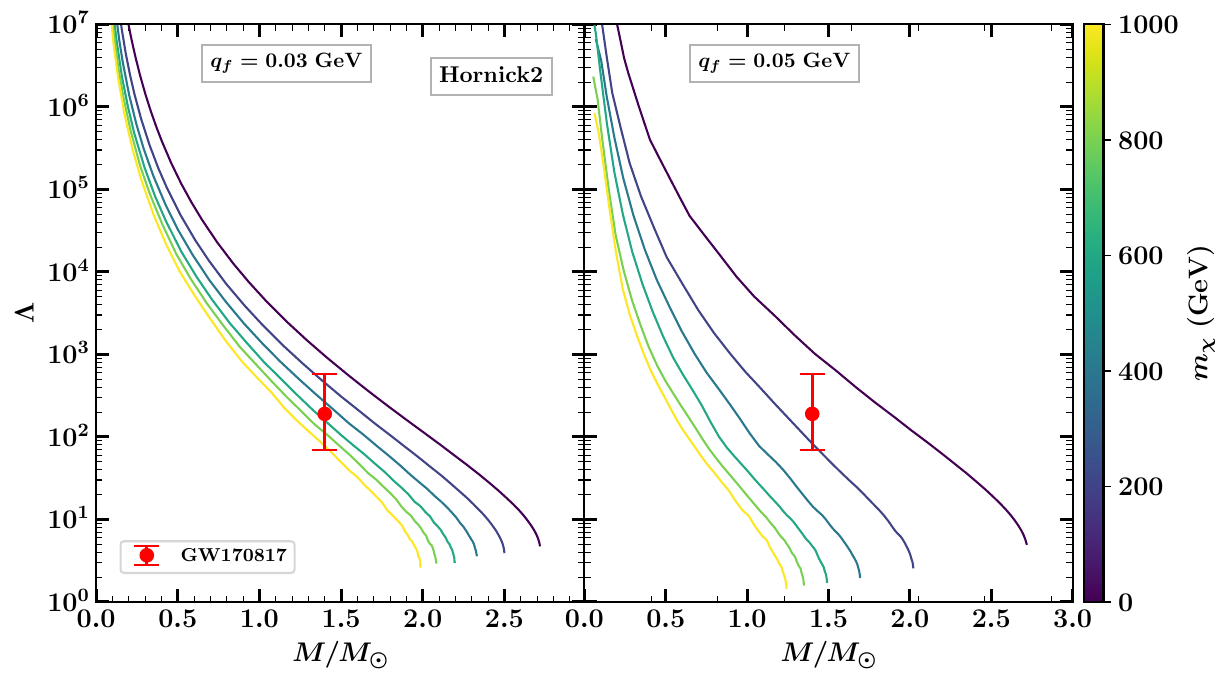}
    \caption{Similar to Fig. \ref{fig:ml_mx}. The DM fermi momentum $q_f = 0.03$ GeV (left) and $q_f = 0.05$ GeV (right) are fixed, with the simultaneous variation of $m_\chi$ from $0$ to $1000$ GeV (color bar).}
    \label{fig:ml_qf}
\end{figure}

Similarly, in Fig. \ref{fig:ml_qf}, the $\Lambda - M$ relation is shown. Here, $q_f$ is kept constant at $0.03$ GeV (left panel) and $0.05$ GeV (right panel). At the lower $q_f$, in the left panel, it is observed that almost all the curves meet the observational bound imposed by the GW170817 event \cite{Abbott_2017}. However, at the higher $q_f$ value in the right panel, the DM becomes more effective, and the degree of reduction of the tidal deformability becomes greater. For this reason, most of the curves seem to deviate from the observational bounds. The tidal constraint therefore excludes high value of both $q_f$ and $m_\chi$. These high values also disfavor the mass-radius constraints, which have been discussed in earlier sections.

The results presented above are derived from the Hornick2 hadronic EOS. Given that the observables of NS are influenced by the fundamental hadronic EOS, it is crucial to investigate whether the impacts caused by ALP-mediated dark matter are significantly reliant on the selection of the hadronic parameterization. We therefore examine the resilience of the ALP-mediated DM effects by including two supplementary hadronic parameterizations, namely NITR-I \cite{pinku_ijmpe_2024} and QMC-RMF4 \cite{Alford_2022}, alongside Hornick2.
\begin{figure}
    \centering
    \includegraphics[width=0.65\textwidth]{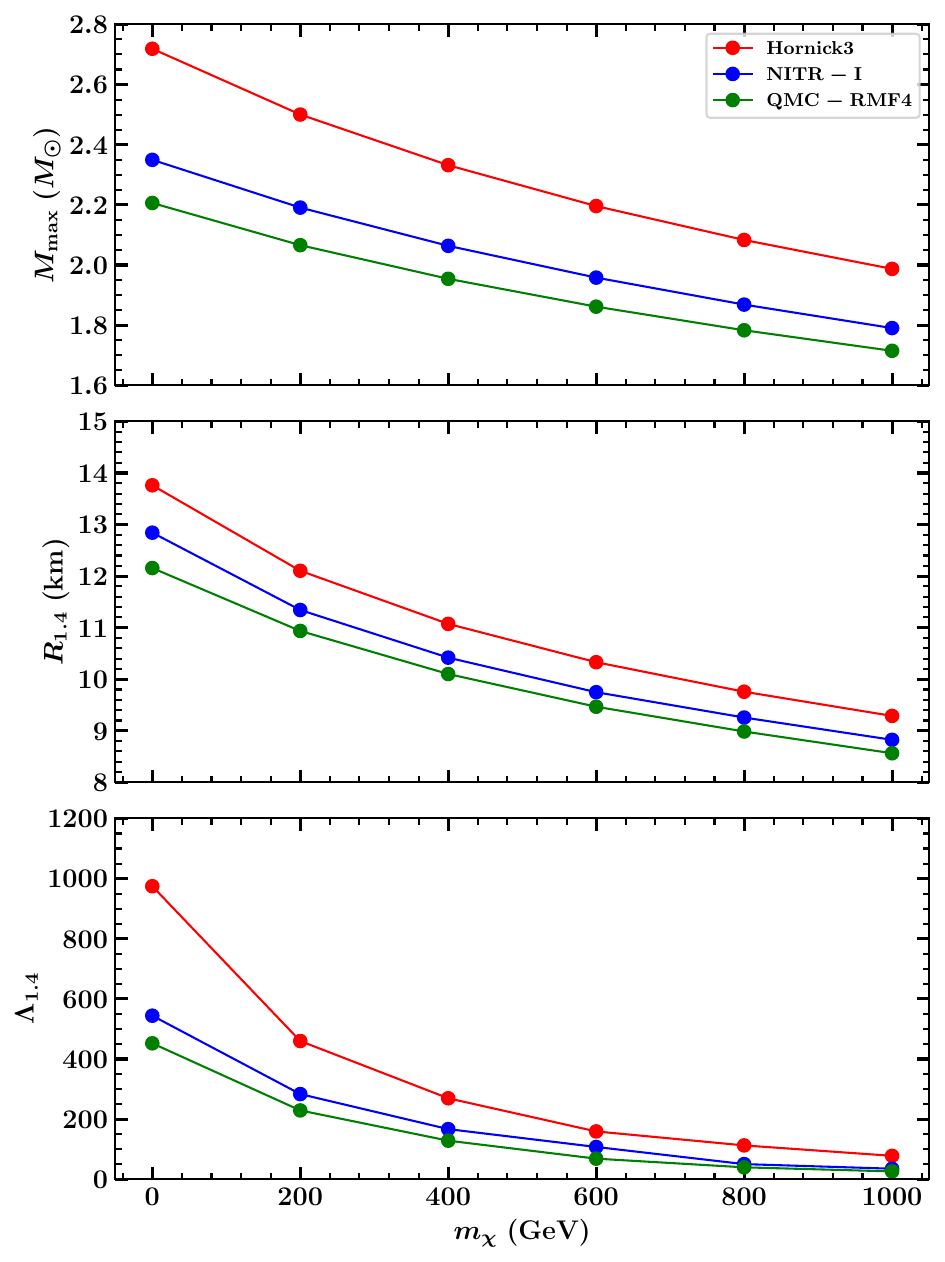}
    \caption{Variation of the maximum mass, canonical radius $R_{1.4}$, and canonical tidal deformability $\Lambda_{1.4}$ with the DM particle mass $m_\chi$ for the Hornick2, NITR-I, and QMC-RMF4 hadronic EOSs, at a fixed DM Fermi momentum of $q_f=0.03$ GeV.}
    \label{fig:mrl}
\end{figure}

The illustration in Fig. \ref{fig:mrl} depicts the fluctuation of the maximum mass $M_{\rm max}$, canonical radius $R_{1.4}$, and canonical tidal deformability $\Lambda_{1.4}$ in relation to $m_\chi$, while maintaining $q_f=0.03$ GeV. The three hadronic parameterizations span a relatively broad range of maximum masses, with $M_{\rm max}=2.71,M_\odot$, $2.35,M_\odot$, and $2.20,M_\odot$ for Hornick2, NITR-I, and QMC-RMF4, respectively. This offers a valuable assessment of the reliance of ALP-mediated DM effects on the stiffness of the underlying hadronic EOS.

As shown in Fig. \ref{fig:mrl}, the incorporation of ALP-mediated DM results in a consistent decrease in the maximum mass, canonical radius, and tidal deformability across all three hadronic EOS assessed. The overall trends with increasing $m_\chi$ are qualitatively aligned across several parameterizations, suggesting that the influence of ALP-mediated DM is not confined to the specific hadronic EOS. However, the magnitude of the changes differs among the three EOSs, reflecting the dependence of the stellar response on the underlying hadronic microphysics. These findings suggest that the qualitative impacts of ALP-mediated DM on NS observables remain consistent regardless of the selected hadronic EOS examined here. Nevertheless, the modifications of $M_{\rm max}$, $R_{1.4}$, and $\Lambda_{1.4}$ are not unique signatures of ALP-mediated DM.

\subsection{Radial oscillation Modes and Stability}  \label{subsec:radial_results}
Radial oscillations furnish a dynamical probe of the stability of the NS. In sec. \ref{subsec:radial}, it was noted that upon resolving the perturbed equations, the squared fundamental radial frequency ($\omega^2$) is derived, signifying stability when it is positive. 

\begin{figure}
    \centering
    \includegraphics[width=1.0\textwidth]{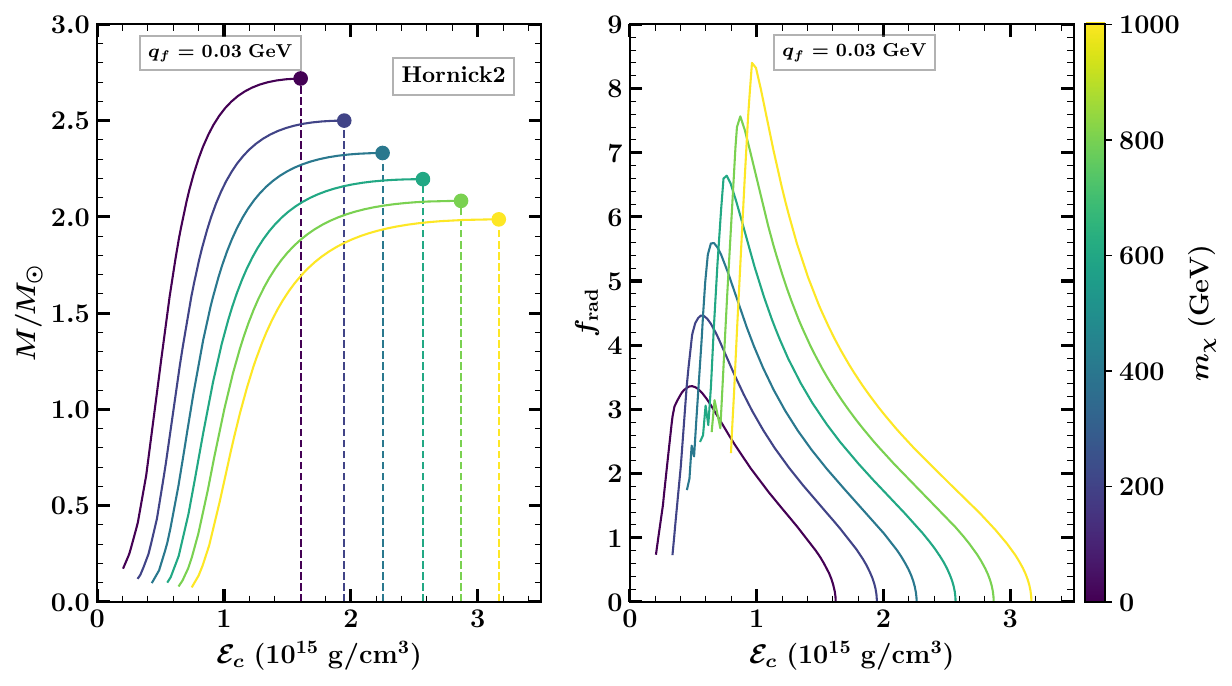}
    \caption{{\it Left:} The gravitational mass is plotted with the central density at fixed $q_f=0.03$ GeV, along with varying $m_\chi$. {\it Right:} Similar to the left panel, but the variation of fundamental radial frequency ($f_{\rm rad}$).}
    \label{fig:rad}
\end{figure}

The mass variation with the central energy density for various $m_\chi$ values at a given $q_f$ is illustrated in the left panel of Fig. \ref{fig:rad}. As discussed in Sec. \ref{subsec:mr}, as $m_\chi$ increases, the maximal gravitational mass is reduced. This trend represents the EOS being gradually softened by the presence of ALP-mediated DM. The maximum mass configurations are shown by the full circles across every curve, while the corresponding vertical lines mark the central energy densities at which these maxima occur. The maximum mass seems to be decreasing as $m_\chi$ increases, but the energy density appears to be shifting to greater values. Such behavior arises due to the inclusion of ALP-mediated DM, which increases the total energy density while providing negligible pressure support. Consequently, a higher central energy density is required to sustain hydrostatic equilibrium, resulting in more compact stellar configurations despite their lower maximum masses.

The fundamental radial frequency ($f_{\rm rad}$) fluctuation with central energy density of the NS is illustrated in the right panel of Fig. \ref{fig:rad}. At a fixed value of $q_f$, the results are analyzed with varying $m_\chi$. For every value of $m_\chi$, the fundamental radial frequency decreases continuously with increasing central energy density and eventually becomes zero at a critical central energy density, $\mathcal{E}_{c,\ crit.}$. This critical point coincides with the central energy density corresponding to the maximum-mass configuration obtained from the $M-\mathcal{E}_c$ relation. In the pure hadronic case ($m\chi=0$ GeV), for instance, the maximum mass is reached at $\mathcal{E}_c=1.60\times10^{15}\ {\rm g/cm^3}$, which is precisely the density at which the fundamental radial frequency vanishes.

The onset of dynamical instability is marked by the vanishing of the squared eigenfrequency of the fundamental radial mode, i.e., $\omega_0^2=0$, at the $\mathcal{E}_{c,\ crit.}$. For $\mathcal{E}_{c} < \mathcal{E}_{c,\ crit.}$, the squared eigenfrequency $\omega_0^2>0$, corresponding to stable oscillatory behavior. Beyond $\mathcal{E}_{c,\ crit.}$, the squared eigenfrequency becomes negative, $\omega_0^2<0$, and the corresponding eigenfrequency becomes imaginary. In this regime, the time dependence of the perturbation changes from oscillatory to exponential, indicating that the stellar configuration is dynamically unstable and cannot restore itself to its equilibrium state following a perturbation. Consequently, the maximum mass configuration represents the last dynamically stable NS, while the configurations that have higher center energy densities tend to be unstable and predicted to experience a gravitational collapse by satisfying the relation $dM/d\mathcal{E}_c>0$. This stability criterion also offers a physical explanation for why the branch of MR-relation beyond the maximum-mass point corresponds to unstable stellar configurations.

Then coming to the effect of the DM on the $f_{\rm rad}$, the presence of ALP-mediated DM modifies the radial oscillation frequency in two important ways. First, at a given central energy density, the magnitude of the fundamental radial frequency increases with increasing $m_\chi$. Second, the $\mathcal{E}_{c, \ crit.}$ shifts systematically to higher values as the $m_\chi$ increases. Specifically, $\mathcal{E}_{c, \ crit.}$ increases from $1.60 \times 10^{15} {\rm g/cm^3}$ for the pure hadronic case to $1.94 \times 10^{15} {\rm g/cm^3}$ and $2.25 \times 10^{15} {\rm g/cm^3}$ for $m_\chi=200$ GeV and $400$ GeV, respectively. This behavior reflects the additional energy density contributed by the DM component, which shifts the onset of dynamical instability to higher central densities even though the maximum gravitational mass decreases.

It is worth noting that the correspondence between the maximum-mass configuration and the onset of radial instability is not universal and depends on the physical assumptions adopted for the stellar equilibrium and perturbations. In particular, Pereira {\it et al.} \cite{Pereira_2018} showed that the usual static stability criterion, $dM/d\mathcal{E}_c>0$, remains valid for hybrid stars with a rapid hadron-quark phase transition, whereas it can break down in the case of slow transitions. For slow phase transitions, the fundamental mode frequency can remain real, indicating dynamical stability, even for some stellar configurations satisfying $dM/d\mathcal{E}_c\leq0$. In a subsequent study, Pereira {\it et al.} \cite{Pereira_2021} investigated the effects of elastic quark phases on the dynamical stability of hybrid stars and found that elasticity can extend the stability window, with the onset of instability occurring at central densities higher than those corresponding to the maximum-mass configurations. Similarly, Canullan-Pascual {\it et al.} \cite{Canullan_Pascual_2025} showed that frozen composition perturbations can allow stable configurations to persist beyond the maximum-mass point. Other studies have demonstrated that additional physical degrees of freedom, including electric charge, pressure anisotropy, and multifluid dynamics, can modify the relation between equilibrium turning points and the onset of radial instability \cite{Arbanil_2015, Kain_2020, Sailesh_radial_2024}.

\subsection{Nonradial oscillation} \label{subsec:nonradial_results}
Nonradial oscillations of the NS present a significant avenue for investigating their internal composition, yielding a diverse range of modes that may interplay with GWs. In this section, we investigate the effects of ALP-mediated DM on the fundamental nonradial frequency ($f_{\rm nrad}$). As described in Sec. \ref{subsec:non_radial}, the nonradial perturbation equations are solved within the full general relativistic formalism.

\begin{figure}
    \centering
    \includegraphics[width=1.0\textwidth]{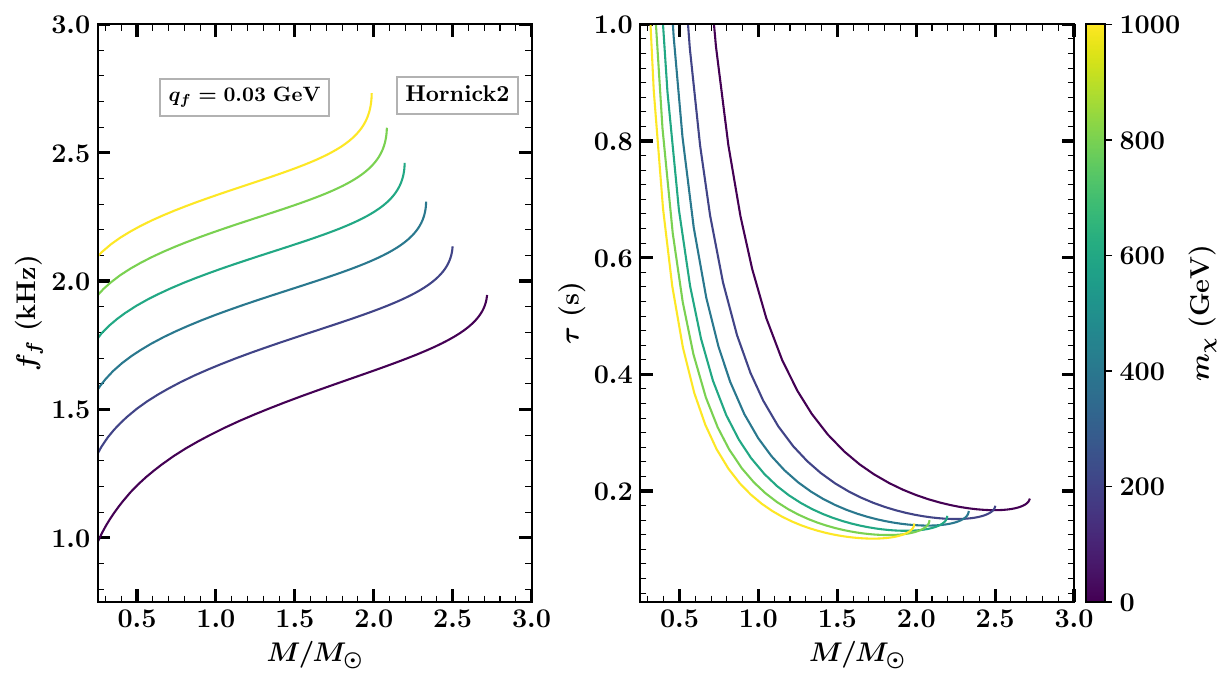}
    \caption{The fundamental nonradial frequency (left) and damping time (right) is shown against the gravitational mass maintaining a constant $q_f = 0.03$ GeV while varying $m_\chi$ as displayed in the color bar.}
    \label{fig:Mfnrad_qf}
\end{figure}

The left panel of fig. \ref{fig:Mfnrad_qf} shows how the fundamental nonradial frequency fluctuates with gravitational mass for different values of the DM parameters, $q_f$ and $m_\chi$. As expected, $f_{\rm nrad}$ increases with stellar mass owing to the corresponding increase in the mean stellar density. Observing the effects of ALP-mediated DM, it systematically shifts the $f_{\rm nrad}$ to higher values for a constant gravitational mass compared to the pure hadronic case ($m_\chi=0$ GeV). Furthermore, the enhancement becomes more pronounced with increasing $m_\chi$, even though the maximum gravitational mass decreases due to the softening in the EOS. For canonical NS, the fundamental nonradial frequencies are $(f_{\rm nrad})_{1.4}=1.51$, $1.75$, $1.95$, $2.12$, $2.27$, and $2.41$ kHz for $m_\chi=0$, $200$, $400$, $600$, $800$, and $1000$ GeV, respectively. These results clearly demonstrate that ALP-mediated DM enhances the fundamental nonradial frequency.

At first glance, the simultaneous decrease in the maximum gravitational mass and increase in $f_{\rm nrad}$ may appear counterintuitive. However, the enhanced compactness induced by DM is mainly responsible for such a phenomenon. As the $m_\chi$ increases, the additional DM energy density compresses the stellar interior, leading to a higher mean density despite the reduction in gravitational mass. Since the $f_{\rm nrad}$ is typically computed by the average density of the star, approximately following $f_{\rm nrad}\propto\sqrt{M/R^{3}}$, the increase in compactness strengthens the restoring force associated with fluid perturbations, resulting in higher oscillation frequencies.

Similarly, in the right panel of Fig. \ref{fig:Mfnrad_qf}, the damping time is shown as a function of the gravitational mass. It is observed that the damping time decreases with increasing stellar mass, which is opposite to the behavior of $f_{\rm nrad}$. This behavior arises because the damping time represents the inverse of the imaginary part of the complex oscillation frequency. Regarding the effect of ALP-mediated DM on the damping time, we observe that the inclusion of DM systematically reduces the damping time. For a canonical NS with $M=1.4 \ M_\odot$, the corresponding values of the damping time are $\tau_{1.4}=0.31$, $0.22$, $0.18$, $0.16$, $0.14$, and $0.13$ s for $m_\chi=0$, $200$, $400$, $600$, $800$, and $1000$ GeV, respectively.

\subsection{Observational prospects and constraints}

The observational implications of ALP-induced DM may be explored via an analysis of NS mass, radius, tidal deformability, and oscillation measurements. Recent X-ray observations, especially from NICER \cite{Miller_2019, Riley_2019, Miller_2021, Riley_2021}, offer progressively accurate limitations on the mass-radius relationship of individual NSs. Similarly, GW observations of binary-NS mergers provide information on the tidal response of the component stars, which can be used, within an assumed theoretical framework, to constrain the properties of dense matter \cite{Abbott_2017}. In the present model, the inclusion of ALP-mediated DM generally reduces the stellar radius and tidal deformability and modifies the maximum mass. These changes indicate that astrophysical observations can be sensitive to the presence and amount of DM in NSs.

Nonetheless, such constraints are not exclusive to the ALP-DM context, since analogous alterations in NS characteristics may also stem from ambiguities in the underlying hadronic EOS, including the appearance of additional degrees of freedom such as hyperons, kaons, $\Delta$-baryons, or quark matter. Consequently, ALP-mediated DM and additional hadronic degrees of freedom can produce partially degenerate effects in macroscopic NS observables. Indeed, previous studies have shown that even a relatively small DM component can modify the interpretation of NS mass, radius, and tidal deformability constraints as probes of the nuclear EOS \cite{Ellis_2018}. Therefore, a measured value of the tidal deformability alone cannot be used to isolate the presence of ALP-mediated DM from uncertainties in the hadronic EOS.

In particular, the tidal deformability inferred from a binary-NS merger is a property of the compact objects entering the GW signal and is interpreted through a model for their masses, tidal responses, and underlying stellar structure. If DM is present inside the stars, it contributes to their equilibrium structure and consequently modifies their tidal response. Thus, applying a standard purely hadronic EOS inference to a DNS can absorb part of the DM effect into the inferred hadronic EOS. Conversely, a change attributed to an unusual hadronic EOS could, in principle, be partly mimicked by a DM component. A reliable assessment of ALP-mediated DM therefore requires a joint treatment of the hadronic EOS and DM parameters rather than using tidal deformability as an independent discriminator.

The large DM fractions obtained in some of the configurations considered in this work should also be interpreted with caution. In our calculations, the DM fraction is specified through the $q_f$ and the resulting equilibrium configurations are studied without modeling the astrophysical mechanism responsible for producing or accumulating the DM component. In particular, we do not calculate DM capture rates, the ambient DM density, the interaction-dependent accumulation history, or the DM content inherited during NS formation. Therefore, the configurations with $f_\chi\gtrsim30\%$ should be regarded as phenomenological equilibrium solutions within the parameter space explored here, rather than as predictions that such large DM fractions are generically realized in astrophysical NSs. The astrophysical viability of large DM fractions is strongly model dependent. Studies of DNS have found that the amount of DM compatible with NS observational constraints can range from relatively small fractions to values above $10\%$, depending on the DM particle properties, self-interaction, and the assumed hadronic EOS \cite{Barbat_2024, Rutherford_2023}. These results emphasize that there is currently no model-independent prediction for the amount of DM that can be present in a NS. In the present work, the configurations with large $f_\chi$ are therefore useful for exploring the structural response of NSs to an appreciable DM component, but their astrophysical realization requires a separate formation and accumulation analysis.

Future observations combining precise mass-radius measurements, binary-NS GW observations, and, potentially, NS oscillation measurements may help reduce these degeneracies. In particular, a joint inference that simultaneously varies the hadronic EOS and the ALP-mediated DM parameters could determine whether a given set of observations is better described by ordinary hadronic physics, by an additional DM component, or by a combination of both. Such an analysis represents an important direction for constraining the ALP-mediated DM scenario.

\subsection{Context with existing NS models}
It is useful to place the present findings within a broader framework of known NS theories congruent with current astronomical observations. A broader range of purely hadronic EOS are able to explain the observed masses and radii of huge NSs \cite{Kumar_2018, Alford_2022, Hornick_2018}. In addition, models incorporating extra degrees of freedom, such as hyperons \cite{Schaffner_hyperon_1996, Weissenborn_hyperon_2012, Bednarek_hyperon_2012, Fortin_hyperon_2015}, quark matter \cite{Weber_quark_1999, Orsaria_quark_2014, Baym_quark_2018, Annala_quark_2020, Ferreira_quark_2020}, or other exotic components such as kaons and $\Delta$-baryons \cite{Thapa_kaon_2020, FuMa_kaon_2022, Probit_2024}, can likewise satisfy the current observational constraints.

Similarly, various studies on the DNS have shown that the presence of DM can modify the mass-radius relation, the tidal deformability, and the oscillation properties, with the magnitude of these effects depending on the DM particle properties, interaction mechanism, and underlying hadronic EOS \cite{Kouvaris_2008, Kouvaris_DM_in_NS_2010, Panotopoulos_2017, Motta_2018, harish_fmode_2021, Sen_2021, Guha_2021, Dutra_2022, Shirke_2024, Divaris_2026}. For instance, WIMP DM can systematically reduce the maximum mass, radius, and tidal deformability of NSs \cite{Panotopoulos_2017, harish_fmode_2021, pinku_prd_2023}. Similarly, NS models involving neutron decay into a dark sector can also lead to reductions in these observables \cite{Motta_2018, Husain_2022, Husain_2023, Shirke_2024, pinku_pdu_2026}.

In the ALP-mediated DM model we consider here, DM usually softens the EOS, leading to a decrease in the maximum mass and radius of the star, and changes the tidal deformability and the frequency of radial and non-radial oscillations. These qualitative tendencies are comparable with the overall behavior reported in numerous DNS models, but the quantitative changes rely on the underlying microphysics. Thus, the alterations seen here should not be taken as unique evidence of ALP-mediated DM, as similar changes in NS observables may also be produced by uncertainties in the hadronic EOS or other kinds of exotic matter. Hence, accurate mass-radius measurements, restrictions on tidal deformability, and future observations of NS oscillations may be necessary to disentangle ALP-mediated DM effects from those associated with other NS models.

\section{Summary and Conclusions}  \label{sec:conclusions}
Through the course of this research, we accomplished an in-depth investigation into how ALP-mediated DM influences the structure, stability, and oscillation properties of NSs. The research was conducted utilizing the Hornick2 EOS, and it was successfully carried out in the context of the RMF formalism. The influence of the DM component was characterized through the $q_f$ and $m_\chi$, allowing us to examine their individual impacts on the global and dynamical properties of NSs.

To begin, we determined the DNS equilibrium configurations by solving the TOV equations. Incorporating ALP-mediated DM gradually softens the EOS by raising the overall energy density with almost minimal pressure increase. As a result, with increasing $q_f$ and $m_\chi$, the star radius and maximum gravitational mass simultaneously diminish. The predicted mass-radius relations contrasted against observational constraints from the pulsars PSR J0740+6620 and PSR J0030+0451, together with the GW event GW170817. We found that NS configurations with relatively small values of $q_f$ and $m_\chi$ remain consistent with current observations, whereas larger values effectively shift the mass-radius curve outside the allowed observational region. The NS observable substantially reduced as the portion of the DM rises with the values of $q_f$ and $m_\chi$, which are described by the parameter $f_\chi$. These results indicate that the parameter space for ALP-mediated DM may be meaningfully constrained using present and future mass-radius measurements. The dimensionless tidal deformability followed the same behavior, where increasing $q_f$ and $m_\chi$ reduces the tidal deformability and further constrains the model through the GW170817 limits.

The dynamical properties of DNS were investigated through both radial and nonradial oscillations. The radial oscillation analysis showed that the fundamental radial mode vanishes accurately at the maximum-mass configuration, confirming that this point marks the onset of dynamical instability. The maximum stable mass is consistently decreased while shifting the critical central energy density to higher values when ALP-mediated DM is included. Consequently, configurations stable in the absence of DM can become dynamically unstable for sufficiently large DM fractions, as indicated by the condition $\omega_0^2<0$. The consistent variations of the maximum stable mass and radial-mode stability features further confirm the structural modifications of NSs related with the EOS softening generated by the ALP-mediated DM.

We also studied the fundamental nonradial mode within the full general relativistic framework. The presence of ALP-mediated DM increases the fundamental frequency for NSs of the same gravitational mass. This behavior originates from the enhanced compactness of the stellar configuration: despite the reduction in gravitational mass, the additional DM energy density compresses the star, increasing its mean density and consequently strengthening the restoring force responsible for the oscillations. In the other hand, the ALP-mediated DM systematically reduces the damping time. It suggests that DNS are likely to oscillate at greater frequencies and undergo increased damping compared to those without DM.

Overall, this work demonstrates that ALP-mediated DM leaves correlated signatures in the NS's equilibrium structure, tidal deformability, radial stability, and nonradial oscillation modes. The combined analysis of these observables offers a cohesive structure to investigate DM-NM interaction in compact stars. Since different mediator portals can modify the NS properties in distinct ways, a natural extension of the present work is to investigate scalar and vector-mediated DM interactions and perform a comprehensive statistical analysis incorporating current and future multimessenger observations in order to tightly limit DM parameter space.

\section*{Acknowledgments}
One of the authors, R. N. Sahoo, gratefully acknowledges financial support from the ANRF, Government of India, through the Ramanujan Fellowship  with Ref. No. - RJF/2025/000314.

\bibliography{main}

\end{document}